\documentclass[aps,twocolumn,superscriptaddress,footinbib,longbibliography]{revtex4-1} 

\usepackage[utf8]{inputenc}
\usepackage{longtable}
\usepackage{placeins}
\usepackage{upgreek, textgreek}
\usepackage[dvips]{graphicx}
\usepackage{xcolor}
\usepackage{epsfig,graphicx,amsfonts,amsbsy}
\usepackage{amsmath,amsfonts,amsthm,amssymb}
\usepackage{appendix}
\usepackage{makeidx}
\usepackage{url}
\usepackage{verbatim}
\usepackage[bookmarksnumbered,pdfpagelabels=true,plainpages=false,colorlinks=true,linkcolor=blue,citecolor=red,urlcolor=blue]{hyperref}
\usepackage[rightcaption]{sidecap}
\usepackage{array}
\usepackage{multirow}
\usepackage{tabularx}
\usepackage{braket} 
\usepackage{dsfont}
\usepackage{bbold}
\usepackage{etoolbox}
\usepackage{diagbox}
\usepackage{float}
\usepackage{dcolumn}
\usepackage{bm}
\usepackage{blindtext}
\usepackage{comment}
\usepackage{physics}

\renewcommand{\r}{\boldsymbol{r}}
\newcommand{\R}{\boldsymbol{R}}
\newcommand{\h}{\boldsymbol{h}}
\newcommand{\A}{\boldsymbol{A}}
\newcommand{\B}{\boldsymbol{B}}
\renewcommand{\j}{\boldsymbol{j}}
\newcommand{\f}{\boldsymbol{f}}

\newcommand{\xhat}{\boldsymbol{\hat{x}}}
\newcommand{\yhat}{\boldsymbol{\hat{y}}}
\newcommand{\zhat}{\boldsymbol{\hat{z}}}
\newcommand{\nhat}{\boldsymbol{\hat{n}}}

\begin{document}

\title{Manipulating Vortices with Domain Walls in Superconductor-Ferromagnet Heterostructures}

\author{Sebasti{\'a}n A. D{\'i}az}
\affiliation{Faculty of Physics and Center for Nanointegration Duisburg-Essen (CENIDE), University of Duisburg-Essen, 47057 Duisburg, Germany}
\affiliation{Department of Physics, University of Konstanz, 78457 Konstanz, Germany}

\author{Jonas Nothhelfer}
\affiliation{Faculty of Physics and Center for Nanointegration Duisburg-Essen (CENIDE), University of Duisburg-Essen, 47057 Duisburg, Germany}

\author{Kjetil M. D. Hals}
\affiliation{Department of Engineering Sciences, University of Agder, NO-4879 Grimstad, Norway}

\author{Karin Everschor-Sitte}
\affiliation{Faculty of Physics and Center for Nanointegration Duisburg-Essen (CENIDE), University of Duisburg-Essen, 47057 Duisburg, Germany}

\date{\today}

\begin{abstract}
Vortices are point-like topological defects in superconductors whose motion dictates superconducting properties and controls device performance.
In superconductor-ferromagnet heterostructures, vortices interact with topological defects in the ferromagnet such as line-like domain walls.
While in previous heterostructure generations, vortex-domain wall interactions were mediated by stray fields; in new heterostructure families, more important become exchange fields and spin-orbit coupling.
However, spin-orbit coupling’s role in vortex-domain wall interactions remains unexplored.
Here we uncover, via numerical simulations and Ginzburg-Landau theory, that Rashba spin-orbit coupling induces magnetoelectric interactions between vortices and domain walls that crucially depend on the wall’s winding direction---its helicity.
The wall’s helicity controls whether vortices are pushed or dragged by Néel walls, and their gliding direction along Bloch walls.
Our work capitalizes on interactions between topological defects from different order parameters and of different dimensionality to engineer enhanced functionality.
\end{abstract}

\maketitle

\section{Introduction}

Nature provides us with numerous examples where nanostructures of dissimilar kind and dimensionality work together to achieve specific functions: 
supramolecular capsules trap small molecules and act as catalytic centers~\cite{Zhang2015,Syntrivanis2020}, lipid bilayer membranes enclose organelles that determine cells’ functionality~\cite{Alberts2022}, RNA polymerase glides along a DNA strand to transcribe genetic information~\cite{Lee2000,Suzuki2012}.
In the spirit of these examples from Nature, topological defects of ordered media in condensed matter~\cite{Mermin1979,ChaikinLubensky2000}, such as point-like skyrmions~\cite{Muhlbauer2009,Yu2010,Bogdanov2003,Everschor-Sitte2018,Shen2021} and vortices~\cite{Hubert1998,Shinjo2000,Blatter1994,Olli1999,Salomaa1987}, as well as line-like dislocations~\cite{Hull2011,Kleman1989,Schoenherr2018,Azhar2022} and domain walls~\cite{Malozemoff1979,Meier2020,Nataf2020,Meier2022}, have been actively investigated for their potential use in technological applications.
For instance, recent experiments have reported on the confinement of skyrmions using domain walls in liquid crystals and magnets~\cite{Foster2019,Tang2021}. 
Similarly, recent theoretical studies~\cite{Knapman2021,Kong2021} and their experimental realization~\cite{Song2022} have demonstrated the guidance of magnetic skyrmions along the tracks provided by the helical phase in helimagnets~\cite{Ezawa2011,Muller2017,Stepanova2022}.
The main challenge faced by these recent reports was ensuring the coexistence of topological defects with different dimensionality and of the same order parameter.

An alternative to overcome this challenge and a path towards engineering enhanced functionality, as in the above examples from Nature, is to employ different order parameters.
Progress in this direction has already been made in superconductor-ferromagnet heterostructures~\cite{Lyuksyutov2005,Buzdin2005,Aladyshkin2009}, where magnetic domain walls have been shown to interact with superconducting vortices via stray fields~\cite{Helseth2002,Vestgarden2007}. 
Moreover, both the displacement~\cite{Goa2003,Vlasko-Vlasov2008} and the confinement~\cite{Iavarone2011} of superconducting vortices by the stray field of magnetic domain walls have already been experimentally demonstrated.

Current interest in proximity-coupled superconductor-magnet heterostructures stems from research on topological superconductivity~\cite{Fu2008,Lutchyn2010,Oreg2010} and the promise of realizing Majorana bound states~\cite{Kitaev2000,Alicea2012} whose manipulation, for example, could be achieved via controlling magnetic skyrmions~\cite{Nothhelferpatent2019, Nothhelfer2019, Diaz2021,Nothhelfer2022}.
Spin-orbit coupling (SOC), necessary for topological superconductivity, is also present in heterostructures hosting superconducting vortices and magnetic domain walls, but its role has so far been neglected and remains unexplored.
\begin{figure}[t]
\centering
\includegraphics[width=0.8\columnwidth]{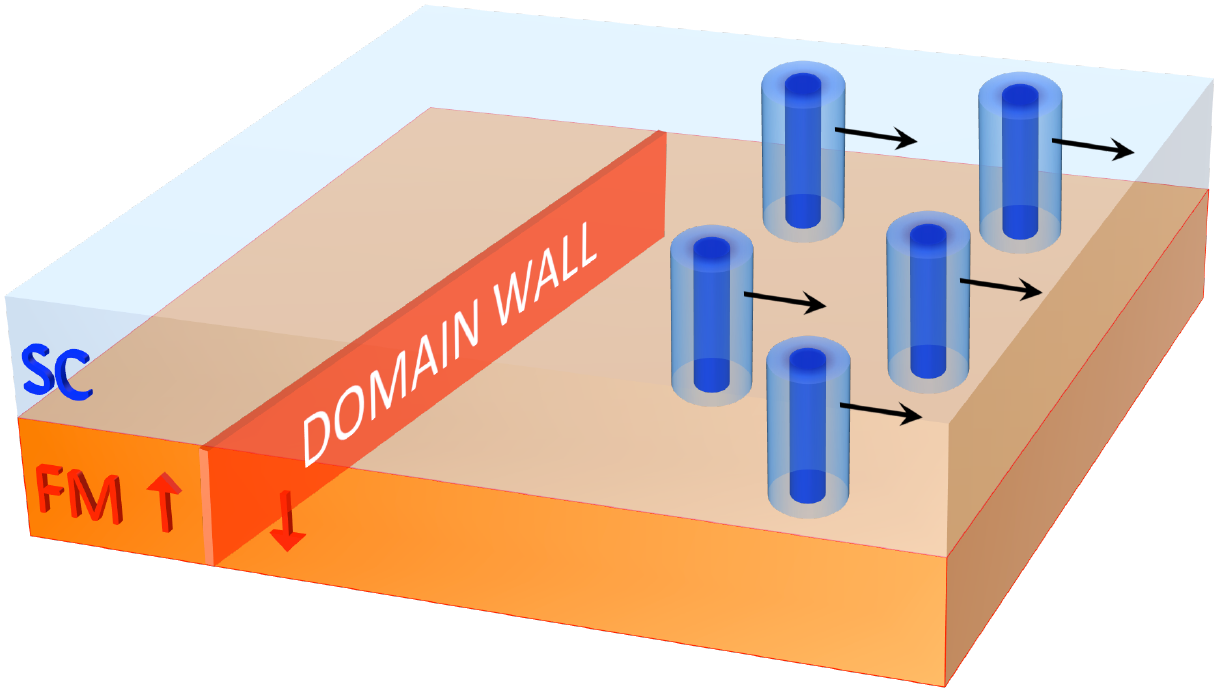}
\vspace{-0.3cm}
\caption{Superconducting vortices manipulated by magnetic domain walls in superconductor-ferromagnet heterostructures. 
A magnetic domain wall in the ferromagnetic thin film (FM) proximity-coupled to a superconducting layer (SC) interacts with superconducting vortices (blue tubes) via the magnetoelectric effect induced by spin-orbit coupling.
The winding of the domain wall (helicity) and the vortices (vorticity) control whether vortices glide along the domain wall, get dragged, or (as depicted) pushed.
}
\label{fig:Setup}
\end{figure}

Here we exploit the rich interplay between the distinct order parameters in proximity-coupled superconductor-ferromagnet heterostructures, arising from the magnetoelectric effect induced by Rashba SOC~\cite{Edelstein1995,Edelstein1996,Yip2002}, to manipulate superconducting vortices with magnetic domain walls.

\section{Textured ferromagnet proximity-coupled to a superconductor}

We assume the magnetic domain wall dynamics to be externally controlled, for instance by in-plane magnetic fields~\cite{Fang1990,Kabanov2010} or temperature gradients~\cite{Berger1985,Torrejon2012,Jiang2013}, and aim to model its effect on the superconductor.
Moreover, the phenomena we are interested in possess a natural separation of time scales.
While magnetic domain walls attain speeds of $\sim 10$~$\mu\rm{m/s}$ when driven by in-plane magnetic fields~\cite{Kabanov2010} and of $\sim 10^2$~$\mu\rm{m/s}$ when driven by temperature gradients~\cite{Jiang2013}, itinerant electrons in clean superconductors are orders of magnitude faster due to typical Fermi velocities of $\sim 10^6$~m/s~\cite{AshcroftMermin1976}.
Therefore, we adopt an adiabatic approach where we simulate the time evolution of the superconducting order parameter as a sequence of static configuration steps.  
At each step, the magnetic texture in the proximity-coupled ferromagnet is a parameter set that modifies the electronic energy spectrum and eigenstates in the superconductor through the tight-binding Hamiltonian
\begin{align}\label{eq:TightBindingHamiltonian}
H=&-t \sum_{\langle ij\rangle} \boldsymbol{c}_i^{\dagger}\boldsymbol{c}_j -\mu \sum_{i}\boldsymbol{c}_i^{\dagger}\boldsymbol{c}_i-\sum_i\boldsymbol{c}_i^{\dagger} (\boldsymbol{h}_i\cdot\boldsymbol{\sigma})\boldsymbol{c}_i \\
&+\mathrm{i} \, \alpha_{\text{R}} \sum_{\langle ij\rangle} \boldsymbol{c}_i^{\dagger}\hat{\boldsymbol{z}}\cdot (\boldsymbol{\hat{d}}_{ij}\times\boldsymbol{\sigma} )\boldsymbol{c}_j +\sum_i (\Delta_ic_{i\uparrow}^{\dagger}c_{i\downarrow}^{\dagger}+h.c.) \,. \nonumber
\end{align}
Itinerant electrons of spin $\alpha$ are created (annihilated) at site $i$ of the square lattice, on the $xy$ plane and with lattice constant $a$, via the operator $c^\dagger_{i\alpha}$($c_{i\alpha}$), and $\boldsymbol{c}_i = (c_{i\uparrow},c_{i\downarrow})^T$. 
Here $t$ is the hopping parameter, $\mu$ is the chemical potential, and $\boldsymbol{h}_i$ is the exchange field due to the magnetic texture in the proximity-coupled ferromagnet.
Rashba SOC enters through the fourth term where $\alpha_{\text{R}}$ is the coupling strength, the unit vector $\boldsymbol{\hat{d}}_{ij}$ is oriented along the nearest-neighbor link from $j$ to $i$, $\boldsymbol{\sigma}$ is the vector of Pauli matrices, and the coefficient $\mathrm{i}$ is the imaginary unit.
Finally, $\Delta_i$ is the complex-valued and spatially-dependent superconducting order parameter.

Since we aim to describe a ferromagnet proximity-coupled to a superconductor whose order parameter is spatially-dependent, a self-consistent treatment is mandatory.
A successfully convergent self-consistent calculation yields the electronic energy spectrum and eigenstates as well as the spatially-dependent superconducting order parameter.
Details of this self-consistent calculation approach can be found in~\cite{Nothhelfer2022}.

With the adiabatic, self-consistent model outlined above at our disposal, we will now characterize the possible interactions between magnetic domain walls and superconducting vortices.

\section{Vortex response to a moving domain wall}

\begin{figure*}[ht]
\centering
\includegraphics[width=\textwidth]{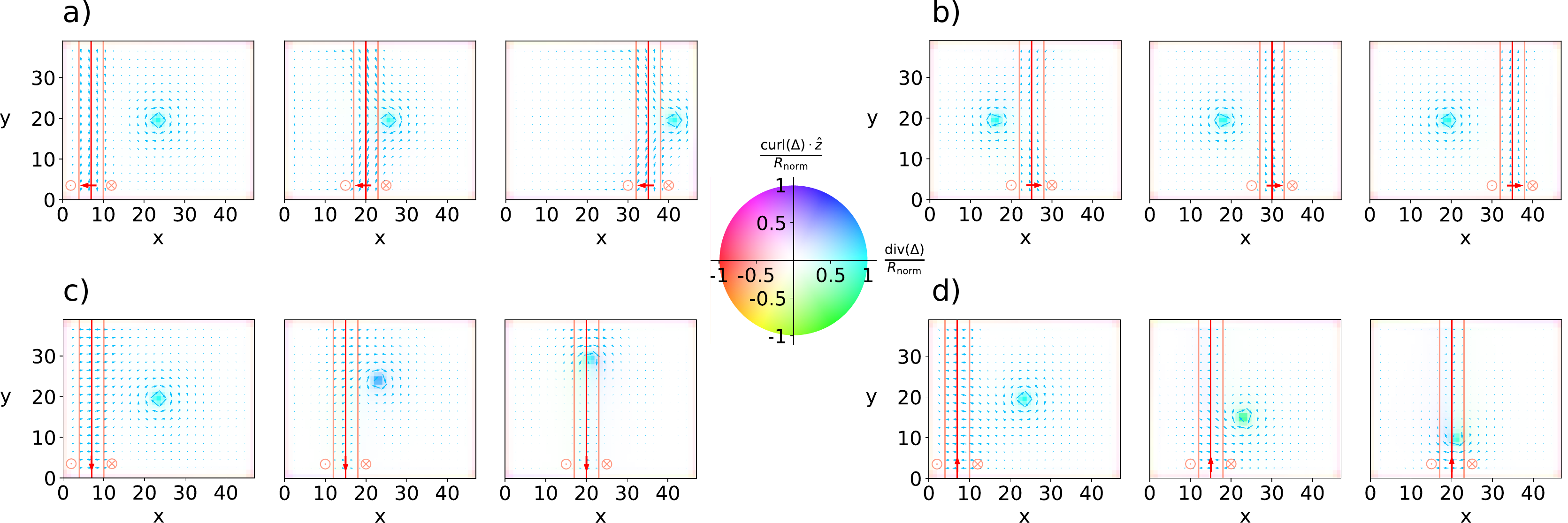}
\caption{
The vortex response to a moving domain wall depends on the wall's helicity.
Snapshots show the time evolution of a vortex in response to a domain wall (vertical red line) moving rightward with helicity (red arrow) (a) $\gamma = \pi$, (b) $\gamma = 0$, (c) $\gamma = 3\pi/2$, and (d) $\gamma = \pi/2$.
Lengths are in units of the lattice constant $a$.
The spatial configuration of the superconducting order parameter is color-coded (central color wheel, normalization constant $R_{\rm{norm}} = 0.75$) by its curl and divergence.
The simulation starts with the vortex on the right domain.
The supercurrent (light blue arrows) is largest in the vicinity of the vortex and of the domain wall.
A N{\'e}el domain wall with (a) $\gamma = \pi$ constantly pushes the vortex. 
In contrast, a N{\'e}el domain wall with (b) $\gamma = 0$ initially attracts the vortex, lets it move to the left domain, and then drags the vortex.
After getting trapped by a Bloch domain wall, the vortex glides along the domain wall in the $\yhat$ direction for (c) $\gamma = 3\pi/2$ and in the $-\yhat$ direction for (d) $\gamma = \pi/2$.
}
\label{fig:DW-SC}
\end{figure*}

To reveal the interaction between a superconducting vortex and a magnetic domain wall we simulate their adiabatic time evolution. 
The exchange field induced by the magnetic domain wall, aligned parallel to the $y$-axis and centered at $x = 0$, is given by
\begin{align}
\h^{(dw)} = h_0 \big[ \sin\Theta(x)\,\nhat^{(dw)}(\gamma) + \cos\Theta(x)\zhat \big] \,, 
\label{eq:h}
\end{align}
where $h_0$ is the exchange field strength, $\nhat^{(dw)}(\gamma) = \cos\gamma\,\xhat + \sin\gamma\,\yhat$, and $\gamma$ is the domain wall \emph{helicity}. 
We choose a wall profile with the magnetic moments pointing asymptotically upward (downward) to the left (right) of the domain wall: $\cos\Theta(x) = -\tanh(x/d)$, where $d$ denotes the domain wall width.

In the vicinity of the vortex, the superconducting order parameter phase adopts the form $\theta^{(v)} = q_v\phi + \varphi$.
The winding number of $\theta^{(v)}$ about the vortex core, its \emph{vorticity} $q_v$, as well as its \emph{chirality} $\varphi$ are determined by the background exchange field at the vortex location and the Rashba SOC strength $\alpha_{\text{R}}$.
For concreteness, we will focus on superconducting vortices with $q_v = 1$~
\footnote{We do so by identifying model parameters that make a $q_v = 1$ vortex, in a field polarized magnetic background, the lowest-energy topologically nontrivial configuration of the superconducting order parameter.}.

We simulate four distinct helicity cases: N{\'e}el domain walls with $\gamma = 0,\pi$ and Bloch domain walls with $\gamma = \pi/2,3\pi/2$ (see App.~\ref{app:vortex_dw_sims}).
Implementing a rigid, rightward domain wall motion yields strikingly different superconducting vortex responses for the four different cases.

A N{\'e}el domain wall with $\gamma = \pi$, when sufficiently close, constantly pushes the superconducting vortex to the right as shown in Fig.~\ref{fig:DW-SC}(a) (see video Neel-Push.mp4). 
However, if the N{\'e}el domain wall has helicity $\gamma = 0$, as Fig.~\ref{fig:DW-SC}(b) depicts, a drastically different behavior occurs. 
The superconducting vortex is attracted towards the domain wall and eventually crosses to the left domain. 
Once in the ``wake'' of the domain wall, the vortex settles at a distance and the domain wall begins to drag the vortex (see video Neel-Drag.mp4).

In contrast, both Bloch domain walls initially attract the superconducting vortex. 
Once sufficiently close, the vortex chirality changes and the vortex motion acquires a vertical component. 
Eventually, the vortex gets trapped and then it glides along the domain wall: for $\gamma = 3\pi/2$ in the $\yhat$ direction, as shown in Fig.~\ref{fig:DW-SC}(c) (see video Bloch-Glide-Up.mp4); and for $\gamma = \pi/2$ in the $-\yhat$ direction, as shown in Fig.~\ref{fig:DW-SC}(d) (see video Bloch-Glide-Down.mp4).

The diverse features of the above four cases admit a unified description underpinned by the magnetoelectric effect induced by Rashba SOC~\cite{Edelstein1995,Edelstein1996,Yip2002}. 
The exchange field due to the magnetic domain wall generates a magnetoelectric supercurrent that interacts with the supercurrent circulating the superconducting vortex, as we describe in the following.

\section{Vortex-domain wall interactions}

The numerical simulations presented above can be comprehended qualitatively through the Ginzburg-Landau theory of superconductors~\cite{Tinkham1996}.
The spatially asymmetric Rashba SOC introduces coupling mechanisms between the superconducting condensate and the ferromagnetic layer, referred to as magnetoelectric interactions~\cite{Edelstein1995,Edelstein1996,Yip2002,Pershoguba2015,Hals2016}.
These interactions are incorporated into the Ginzburg-Landau phenomenology through the following magnetoelectric free energy functional
$F_{me}\left[\psi^{\ast},\psi, \A, \h \right]=  \int d\r \, \mathcal{F}_{me}^{(1)}$, where the free energy density $\mathcal{F}_{me}^{(1)}$ is a SOC-induced Lifshitz invariant given by 
\begin{align}
\mathcal{F}_{me}^{(1)} = -\kappa^{(1)} (\zhat\times\h)\cdot \boldsymbol{\mathcal{P}} \,,
\end{align}
and the positive constant $\kappa^{(1)}$ parameterizes the magnetoelectric coupling to first order in the SOC~\footnote{The contribution to the magnetoelectric coupling to second order in the SOC, as our calculations in App.~\ref{app:js_plots-kappa2} show, is negligible.}.
Here, $\h$ is the exchange field induced in the superconductor by the adjacent ferromagnetic layer via the exchange proximity effect, $\boldsymbol{\mathcal{P}} =  \Re \left[ \psi^{\ast}  (-\mathrm{i}\hbar \boldsymbol{\nabla} - 2e \A/c  ) \psi  \right]$ is the momentum density of the superconducting condensate, $\psi (\r) = | \psi (\r) | \exp (\mathrm{i} \theta (\r) ) $ is the complex order parameter field of the superconductor, $2e$ (where $e = - |e|$) is the charge of a Cooper pair, $c$ is the speed of light, and $\A$ is the magnetic vector potential.
We concentrate on the case of a thin film type-II superconductor where the penetration depth of the magnetic field is significantly larger than the typical length scales that characterize the spatial variations of magnetic domain walls and superconducting vortices. 
In this case, we can disregard the magnetic vector potential~\cite{Kopnin2001b}.
The magnetoelectric free energy functional $F_{me}$ establishes a connection between the orientation of the ferromagnet's magnetization and the momentum density of the condensate. 
An essential consequence of this interaction is the emergence of an anomalous supercurrent density given by (see App.~\ref{app:GL_theory})
\begin{equation}
\j_{as} = - 2e |\psi|^2 \kappa^{(1)} (\zhat\times\h ) \,.
\label{eq:jsA}
\end{equation} 
This is an additional contribution to the conventional supercurrent $\j_{cs}= e\boldsymbol{\mathcal{P}}/m$ generated by phase gradients in $\psi$, resulting in a net supercurrent density of $\j_s= \j_{cs} + \j_{as}$.
Therefore, in proximity-coupled superconductor-ferromagnet heterostructures, textures with an in-plane component such as ferromagnetic domain walls, according to Eq.~\eqref{eq:jsA}, produce anomalous supercurrents.

Incorporating the exchange field defined in Eq.~\eqref{eq:h}, we find the following anomalous supercurrent density associated with a domain wall of general helicity
\begin{equation}
\j_{as}^{(dw)} (x,\gamma) = - g^{(1)} (x)\; \zhat \times \nhat^{(dw)}(\gamma) \,, 
\label{eq:jsDW}
\end{equation} 
where $g^{(1)}(x) = \kappa^{(1)} 2e h_0 |\psi|^2 \sech \left( x/d \right)$ represents the current originating from $\mathcal{F}_{me}^{(1)}$.
While N\'eel domain walls generate supercurrents along the wall, Bloch domain walls generate supercurrents across the wall.

On the other hand, a superconducting vortex centered at $\R = R_x\,\xhat + R_y\,\yhat$ can be characterized by the order parameter field~\cite{Tinkham1996}
\begin{equation}
\psi^{(v)}(\r) = \psi_0 \tanh ( |\r-\R|/\xi ) \exp (\mathrm{i} \theta^{(v)} (\r)) \,, 
\label{eq:PsiV}
\end{equation}
where $\theta^{(v)}$ varies by $2 \pi$ in making a complete circuit counterclockwise about the center of the vortex, $\psi_0$ denotes the absolute value of the order parameter far from the vortex core, and $\xi$ represents the coherence length. 
This vortex profile generates the conventional supercurrent density
\begin{equation}
\j_{cs}^{ (v)} = \frac{\hbar e \psi_0^2}{m} \frac{\tanh^2 (|\r- \R |/\xi)}{|\r- \R |^2} \zhat\times (\r - \R ) \,, 
\label{eq:jsV}
\end{equation} 
which is strongest when $|\r-\R| \sim \xi$ and approaches zero in the limits $|\r-\R| \rightarrow 0$ and $|\r-\R| \rightarrow \infty$.

The vortex-domain wall magnetoelectric interactions are governed by the effective free energy 
\begin{equation}
F_{\rm eff}(\R, \gamma) = F_{me}[\psi^{(v)\ast}, \psi^{(v)}, \h^{(dw)}] \,, \label{eq:Feff}
\end{equation}
which is obtained by substituting $\h^{(dw)}$, the exchange field of a domain wall with helicity $\gamma$ centered at $x=0$ Eq.~\eqref{eq:h}, and $\psi^{(v)}$, the profile of a vortex centered at $\R$ Eq.~\eqref{eq:PsiV}, into the magnetoelectric energy functional $F_{me}$, followed by integrating over the spatial coordinate $\r$. 
Through its dependence on $\R$ and $\gamma$, $F_{\rm eff}$ determines how the superconductor's free energy varies with the vortex's position relative to the domain wall.
Additionally, from Eq.~\eqref{eq:Feff} we can deduce the magnetoelectric force $\f_{me}= - \partial F_{\rm eff}/\partial \R$, which drives the vortex towards the effective free energy minimum.

\begin{figure}[th] 
\centering 
\includegraphics[scale=1.0]{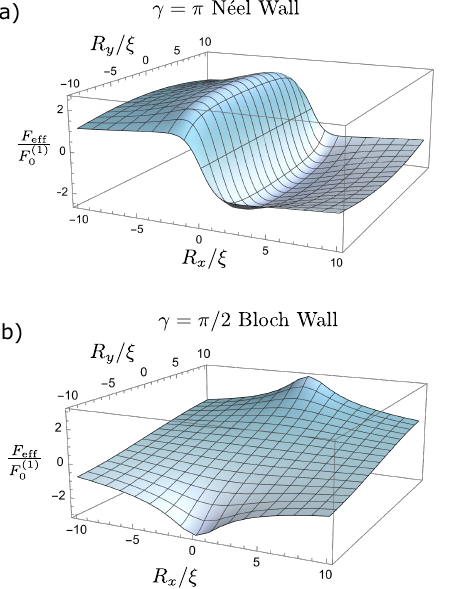} 
\caption{
Effective free energy landscapes of a vortex interacting with a domain wall.
The minima of $F_{\rm eff}$ Eq.~\eqref{eq:Feff}, with the domain wall fixed at $R_x = 0$, indicates the vortex equilibrium position.
The numerical calculation considered a sample of size $[-\infty, +\infty] \times [-10\,\xi, 10\,\xi]$, the domain wall width is $d = 0.4\,\xi$, and $F_0^{(1)} = \kappa^{(1)} \pi \hbar \, d \, h_0 \psi_0^2$.
(a) For a N\'eel wall with helicity $\gamma = \pi$, $F_{\rm eff}$ attains its global minimum to the right of the wall's center at $(R_x \sim \xi, R_y=0)$; this explains the simulation [Fig.~\ref{fig:DW-SC}(a)] where the wall pushes the vortex.
(b) For a Bloch wall with helicity $\gamma = \pi/2$, $F_{\rm eff}$ attains its global minimum at one of the wall ends on the sample boundary at $(R_x = 0, R_y = -10\,\xi)$; this explains the simulation [Fig.~\ref{fig:DW-SC}(d)] where the vortex glides along the wall.
}
\label{fig:Fme} 
\end{figure}

We numerically compute the effective free energy, assuming 
the domain wall width and the coherence length are similar $d \sim \xi$,
in the vicinity of N\'eel [Fig.~\ref{fig:Fme}(a)] and Bloch [Fig.~\ref{fig:Fme}(b)] domain walls.
For $\gamma = \pi$ ($\gamma = 0$) N\'eel walls, $F_{\rm eff}$ exhibits a global minimum at $R_x \sim \xi$ ($R_x \sim -\xi$) from the wall's center. 
This explains why in the simulation for $\gamma = \pi$ [Fig.~\ref{fig:DW-SC}(a)] the vortex is pushed in front of the domain wall as it moves to the right, while for $\gamma = 0$ [Fig.~\ref{fig:DW-SC}(b)] the vortex crosses into the left domain and is dragged behind the moving wall.
In contrast, for $\gamma = \pi/2$ ($\gamma = 3\pi/2$) Bloch walls, $F_{\rm eff}$ exhibits a global minimum at the end of the wall towards $-\yhat$ ($+\yhat$) where it intersects the sample boundary.
Therefore, the effective free energy gives rise to a $y$-component in the vortex motion [Figs.~\ref{fig:DW-SC}(c),(d)], which explains its gliding along the wall.

Our analytical calculations (see App.~\ref{app:eff_free_energy}) corroborate these numerical results and their helicity-dependence.
For Bloch walls, they also clearly show the removal of the minima and maxima of $F_{\rm eff}$ as the sample size approaches infinity---these minima and maxima are finite-size effects.

A simple physical picture of vortex-domain wall interactions, which complements the above free energy calculations, follows from the interplay between anomalous and conventional supercurrents densities.
When the superconductor-ferromagnet heterostructure contains both a domain wall and a vortex, the superconductor aims to minimize its free energy by relocating the vortex to reduce the total supercurrent density $\j_s= \j_{cs}^{(v)} + \j_{as}^{(dw)}$ to a minimum (see App.~\ref{app:GL_theory}).
In the case of a N\'eel wall, the vortex reaches an equilibrium position at a distance $|R_x| \sim \xi$ from the wall's center because this ensures that the area with the strongest conventional supercurrent density around the vortex partly offsets the anomalous supercurrent density induced along the wall.  
While for Bloch walls, the vortex attains its equilibrium position at one of the wall ends to counteract the anomalous supercurrent density induced in the sample interior.

The detailed understanding of the helicity-dependent vortex-domain wall interactions obtained thus far will be exploited in the following section where we demonstrate practical routes to manipulate superconducting vortices with magnetic domain walls.

\section{Manipulating vortices with domain walls}

Vortex manipulation is relevant for technological applications.
For instance, directional guidance of vortices underpins vortex-based logic platforms~\cite{Nakajima1976,OlsonReichhardt2004,Nothhelfer2022,Keren2023}.
Additionally, confining vortices to narrow constrictions enhances vortex pinning, which extends the magnetic field range of the zero-resistance state~\cite{Cordoba2013} and improves the performance of superconducting electronic devices~\cite{Dobrovolskiy2017}.
We employ the helicity-dependent vortex-domain wall interactions detailed in the previous sections to manipulate vortices with domain walls.

\begin{figure}[t]
\centering
\includegraphics[width=\columnwidth]{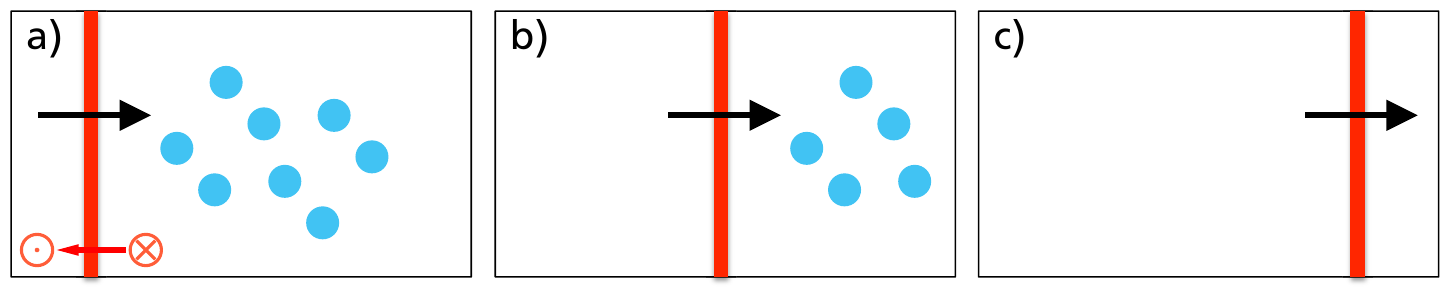}
\caption{Wiping the slate clean of superconducting vortices.
Configurational sequence of a rightward-moving N{\'e}el domain wall with helicity $\gamma = \pi$ that pushes vortices away from the sample area, effectively acting as a vortex ``eraser'' or ``rake'' (see video Vortex-Eraser.mp4).}
\label{fig:Cleaning_Slate}
\end{figure}

A N{\'e}el wall with helicity $\gamma = \pi$ can be utilized to push away and thus remove vortices from a specific sample area, as illustrated in Fig.~\ref{fig:Cleaning_Slate}.
Alternatively, by the same mechanism, an empty sample area can be populated with vortices.
Being mobile barriers, these domain walls also have the potential to suppress or trigger superconducting vortex avalanches ~\cite{Field1995,Altshuler2004,Laviano2017}.

\begin{figure}[t]
\centering
\includegraphics[width=\columnwidth]{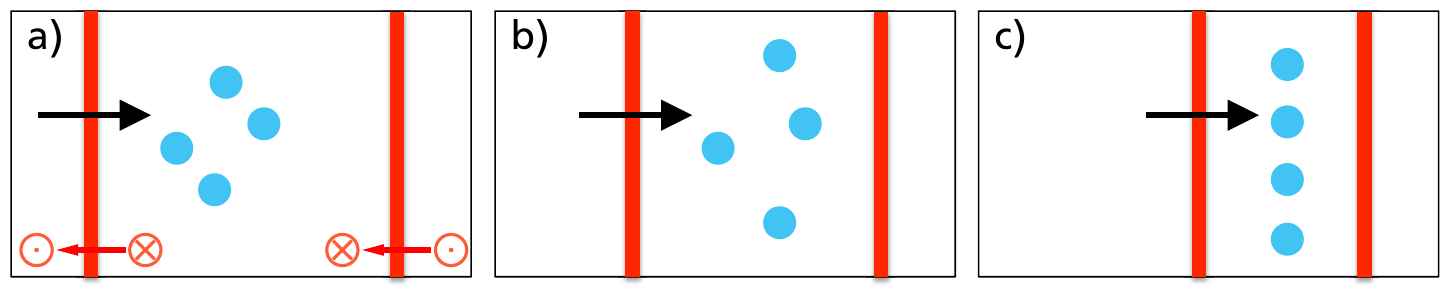}
\caption{Assembling a superconducting vortex chain using two magnetic domain walls.
Configurational sequence of superconducting vortices confined between two N{\'e}el domain walls with helicity $\gamma = \pi$ (see video Vortex-Chain-Assembly.mp4).
As the separation between the walls decreases, the vortices rearrange (due to domain wall pushing and repulsive vortex-vortex interactions) and form a chain parallel to the walls.}
\label{fig:SC_Vortex_Chain}
\end{figure}

Bloch walls themselves, in finite-sized setups, can serve as unidirectional vortex channels where the wall's helicity controls the vortices propagation direction [see Figs.~\ref{fig:DW-SC}(c),(d)].
Another route to construct a vortex channel is depicted in Fig.~\ref{fig:SC_Vortex_Chain} (for a simulation see video Vortex-Chain-Assembly.mp4) where two N{\'e}el walls are employed, also taking advantage of repulsive vortex-vortex interactions, to assemble a chain of vortices.
Although vortex chains have been reported before in anisotropic~\cite{Matsuda2001,Grigorenko2001,Bending2005} and nanopatterned superconductors~\cite{Dobrovolskiy2017,Cordoba2019}, and between twin boundaries~\cite{Song2023}; 
mobile magnetic domain walls represent an attractive non-invasive avenue to construct reconfigurable channels for vortices and vortex chains.

Our simulations demonstrate that exploiting the dependence on the wall's helicity of the vortex-domain wall interactions leads to enhanced functionality.

\section{Discussion and Conclusion}

The experimental realization of our predictions requires superconductor-ferromagnet heterostructures in which vortices and domain walls interact via Rashba SOC.
The high-quality interfaces needed for Rashba SOC can be assembled using van der Waals (vdW) materials.
For example, evidence of Rashba SOC has been reported at the interface of: a quasi-2D layer of the vdW ferromagnet Fe$_{0.29}$TaS$_2$ with NbN~\cite{Cai2021}, and a monolayer of the vdW ferromagnet CrBr$_3$ with NbSe$_2$~\cite{Kezilebieke2020}. 
Both NbN and NbSe$_2$ are conventional, 
type-II superconductors which admit vortices.
Furthermore, it has been established that the CrBr$_3$/NbSe$_2$ heterostructure supports vortices~\cite{Kezilebieke2022} and CrBr$_3$ hosts Bloch domain walls~\cite{Matricardi1967,Sun2021}.
Gliding vortices along Bloch walls [Figs.~\ref{fig:DW-SC}(c),(d)] we propose to observe in the existing CrBr$_3$/NbSe$_2$ heterostructure.
Another vdW ferromagnet that exhibits Bloch domain walls is Fe$_3$GeTe$_2$~\cite{Yang2022}; but by decreasing its thickness or engineering its capping layer, Fe$_3$GeTe$_2$ can favor N{\'e}el domain walls~\cite{Birch2022,Peng2021}.
Vortex pushing or dragging by N{\'e}el walls [Figs.~\ref{fig:DW-SC}(a),(b)] we suggest to observe in Fe$_3$GeTe$_2$/NbSe$_2$ heterostructures.
Therefore, superconductor-ferromagnet heterostructures suitable for the experimental realization of our predictions are available.

Conclusive experimental confirmation of our predictions calls for tracking domain wall and vortex positions.
The observation of domain walls and the determination of their helicity requires the measurement of the three components of the magnetization, which can be achieved using spin-polarized scanning tunneling microscopy (SP-STM)~\cite{Bode2007}, spin-polarized low-energy electron microscopy (SPLEEM)~\cite{Chen2015}, or Lorentz transmission electron microscopy (TEM) even without tilting the sample~\cite{Chess2017}.
Additionally, note that our findings of helicity-dependent vortex-domain wall interactions provide an indirect method to measure the domain wall helicity.
Vortex detection can be done via Lorentz TEM~\cite{Harada1992}, magneto-optical imaging~\cite{Goa2003,Vestgarden2007}, magnetic force microscopy (MFM)~\cite{Moser1995,Auslaender2009}, STM~\cite{Hess1989}, and superconducting quantum interference device (SQUID) magnetometry~\cite{Embon2017}.
Employing a single technique to image vortices and domain walls simultaneously, facilitates data collection and minimizes system disruption.
The simultaneous measurement of domain walls and vortices has been demonstrated using MFM~\cite{Iavarone2011}.
There exist plenty of imaging techniques that allow to confirm our predictions.

We have explained how Rashba spin-orbit coupling in superconductor-ferromagnet heterostructures induces interactions between superconducting vortices and magnetic domain walls.
Leveraging the dependence of these interactions on the domain walls' helicity, we have showcased the enhanced functionalities that result from manipulating vortices with domain walls. 
Our work epitomizes Nature's paradigm of functionality via the interaction of dissimilar nanostructures, and embraces its extension to the condensed matter realm of topological defects of different order parameters and dimensionality.

\begin{acknowledgements}

Jonas Nothhelfer and Sebasti{\'a}n A. D{\'i}az contributed equally to this work.
We thank Lucas G\"orzen for assistance in this work's preliminary stages.
We thank Stephan Kessler, Christos Panagopoulos, Danilo Nikolic, Michael Hein, and Wolfgang Belzig for helpful discussions. 
S. A. D{\'i}az acknowledges funding by a Blue Sky project led by W. Belzig in the framework of the Excellence Strategy of the University of Konstanz.
We acknowledge funding from the German Research Foundation (DFG) Project No. 320163632 (Emmy Noether), Project No. 403233384 (SPP2137 Skyrmionics), the Research Council of Norway via Project No. 334202, and support from the Max Planck Graduate Center. This research was supported in part by the National Science Foundation under Grant No. NSF PHY-1748958.

\end{acknowledgements}


\appendix


\section{Vortex and domain wall simulations}\label{app:vortex_dw_sims}

In the simulations presented in this work, we numerically computed the superconducting order parameter $\Delta$ following the self-consistent calculation detailed in our previous work~\cite{Nothhelfer2022}, and we used the system parameters shown in Table~\ref{tab:parameters}. 
The domain walls, on the other hand, were inputs for the numerical calculation of $\Delta$; as described in Sec.~III in the main text, we assumed the domain wall profile $\cos\Theta(x) = -\tanh(x/d)$.
Here, $d$ denotes the domain wall width, and in our simulations we used $d = 3a$, where $a$ is the lattice constant of the tight-binding Hamiltonian in Eq.~\eqref{eq:TightBindingHamiltonian} of the main text.

Figure~\ref{fig:DW-SC} in the main text shows selected snapshots of four simulations whose videos (Neel-Push.mp4, Neel-Drag.mp4, Bloch-Glide-Up.mp4, Bloch-Glide-Down.mp4) are included in the Supplemental Material. 
In these four simulations, a single vortex is initially placed sufficiently to the right of the domain wall (at a distance of $19$ lattice sites) where the magnetic texture is effectively field polarized.

Also included in the Supplemental Material are simulation videos of the vortex manipulations introduced in Figs.~\ref{fig:Cleaning_Slate}-\ref{fig:SC_Vortex_Chain}: Vortex-Eraser.mp4 and Vortex-Chain-Assembly.mp4.

To obtain the initial vortex configuration of the superconducting order parameter we use a seed at the beginning of the self-consistent calculation.
Subsequent simulation steps use the vortex configuration from the previous step as their seed (for details see our previous work~\cite{Nothhelfer2022}).

\begin{table}[htb]
    \centering
	\begin{tabular}{c|c|c}
		Parameters & Symbol & Value ($t$) \\
		\hline
		Chemical potential & $\mu$ & $-4$\\
		Spin-Orbit-Coupling & $\alpha_{\text{R}}$ & 0.75\\
		Thermal energy & $k_B T$ & 0.001\\
		Debye frequency & $\omega_{D}$ & 100 \\
		Effective attractive interaction & $V$ & 5\\
	\end{tabular}
\caption{System parameters used in this work in units of the hopping $t$.}
\label{tab:parameters}
\end{table}

\section{Ginzburg-Landau theory of noncentrosymmetric superconductors}\label{app:GL_theory}

For noncentrosymmetric superconductors, the Ginzburg-Landau free energy takes the form
\begin{equation}
F \left[\psi^{\ast},\psi, \A, \h \right] =  \int  d\r \left[\mathcal{F}_{0} + \mathcal{F}_{me}^{(1)} + \mathcal{F}_{me}^{(2)}  \right] , \label{eq:GL}
\end{equation}
where $\mathcal{F}_{0}$ is given by
\begin{equation}
\mathcal{F}_{0} = \frac{1}{4m} (\boldsymbol{\Pi} \psi )^{\ast}\cdot \boldsymbol{\Pi} \psi + a |\psi |^2 + \frac{b}{2} |\psi |^4 + \frac{|\B|^2}{8\pi}.
\end{equation}
Here, $\boldsymbol{\Pi} = -\mathrm{i}\hbar \boldsymbol{\nabla} - 2e \A/c  $, and $a$ and $b$ determine the absolute value of the order parameter field $\psi$.
Note that $\psi$ and $\Delta$ are proportional to each other~\cite{Edelstein1996}.
The magnetoelectric free energy densities $\mathcal{F}_{me}^{(1)}$ and $\mathcal{F}_{me}^{(2)}$ are 
\begin{align}
\mathcal{F}_{me}^{(1)} &= - \kappa^{(1)} (\zhat\times\h)\cdot \boldsymbol{\mathcal{P}} \,, \\
\mathcal{F}_{me}^{(2)} &= \kappa^{(2)} (\zhat\times\boldsymbol{\nabla}h_z )\cdot \boldsymbol{\mathcal{P}} \,,
\end{align}
with $\boldsymbol{\mathcal{P}} =  \Re \left[ \psi^{\ast}  (-\mathrm{i}\hbar \boldsymbol{\nabla} - 2e \A/c  ) \psi  \right]$, while the constants $\kappa^{(1)}$ and $\kappa^{(2)}$ parameterize the magnetoelectric coupling respectively to first and second order in the SOC.
Varying the free energy functional in Eq.~\eqref{eq:GL} with respect to $\A$ and using that $\j_s = (c/4\pi) (\boldsymbol{\nabla}\times \B)$ lead to the supercurrent density $\j_s= \j_{cs} + \j_{as}$, where $\j_{cs}= e\boldsymbol{\mathcal{P}}/m$ and, disregarding $\A$ as justified in the main text,
\begin{align}
\j_{as} = 2e |\psi|^2 \left[ - \kappa^{(1)} (\zhat\times\h ) + \kappa^{(2)} (\zhat\times\boldsymbol{\nabla}h_z ) \right] \,.
\label{eq:app_jsA}
\end{align}
Plugging in the domain wall exchange field (Eq.~\eqref{eq:h} in the main text) yields
\begin{align}
\j_{as}^{(dw)} (x,\gamma) = - g^{(1)} (x)\; \zhat \times \nhat^{(dw)}(\gamma) - g^{(2)} (x)\; \yhat \,, 
\label{eq:app_jsDW}
\end{align} 
where $g^{(1)}(x) = \kappa^{(1)} 2e h_0 |\psi|^2 \sech \left( x/d \right)$ and $g^{(2)}(x) = \kappa^{(2)} 2e h_0 |\psi|^2 \sech^2(x/d)/d$. 
The contribution of the term proportional to $\kappa^{(2)}$, as we show in App.~\ref{app:js_plots-kappa2}, is negligible. For this reason, in the main text, this term is absent.

Neglecting $\kappa^{(2)}$ in $\j_{as}$ [Eq.~\eqref{eq:app_jsA}] and $\A$ in $\j_{cs}$, as in the main text, allows the rewrite $\mathcal{F}_{me}^{(1)} = k \, \j_{as} \cdot \j_{cs}$ with $k = m/(2 e^2 |\psi|^2) > 0$, shedding light to our results. 
The superconducting vortex, to minimize $\mathcal{F}_{me}^{(1)}$, places itself in the region near the domain wall where $\j_{as}$ and $\j_{cs}$ are antiparallel and the strongest.

\section{Effective free energy}\label{app:eff_free_energy}

To derive approximate analytical expressions for the effective free energy $F_{\rm eff}$, Eq.~\eqref{eq:Feff} in the main text, we consider a sample that extends infinitely along the $x$-axis and has a width of $2L_y$ along the $y$-axis. 
A domain wall is positioned at $x=0$. 
We assume that the vortex is located near the domain wall and the origin of the sample, satisfying $\tilde{\R} = \R/\xi\ll 1$.
Additionally, we examine narrow domain walls where $d \rightarrow 0$.
By performing a series expansion of $F_{\rm eff}$ in terms of $\tilde{\R}$ to third order and considering the scenarios where the heterostructure hosts N\'eel and Bloch walls, we obtain the following expressions for the effective free energies
\begin{align}\label{eq:FeffNWBW}
F_{\rm eff}^{(nw)} =& (\epsilon F_0^{(1)} + F_0^{(2)}) \left[ c_1 \tilde{R}_x - c_2 \tilde{R}_x^3 - c_3 \tilde{R}_x \tilde{R}_y^2  \right] \,, \\
F_{\rm eff}^{(bw)} =& \epsilon F_0^{(1)} \left[ c_4 \tilde{R}_y + c_5 \tilde{R}_y^3 - c_6 \tilde{R}_x^2 \tilde{R}_y \right] \nonumber\\
& - F_0^{(2)} \left[ c_2 \tilde{R}_x^3 -c_1 \tilde{R}_x  + c_3 \tilde{R}_x \tilde{R}_y^2  \right] \,. \nonumber
\end{align}
Here, $F_{\rm eff}^{(nw)} = F_{\rm eff}|_{\gamma = (1 - \epsilon)\pi/2}$ and $F_{\rm eff}^{(bw)} = F_{\rm eff}|_{\gamma = (2 - \epsilon)\pi/2}$ respectively represent the effective free energies for systems containing N\'eel and Bloch walls, where $\epsilon\in \{ 1, -1 \}$ depending on the domain wall's helicity.
We have also introduced 
$F_0^{(1)} = \pi \hbar d \kappa^{(1)} h_0 \psi_0^2$ and 
$F_0^{(2)} = 2\hbar \kappa^{(2)} h_0 \psi_0^2$. 
In deriving the above expressions, we have utilized the approximations $(1/\pi d) \sech (x/d) \approx \delta (x)$ and $(1/2d) \sech^2 (x/d) \approx \delta (x)$ in the limit $d \rightarrow 0$ (where $\delta(x)$ is the Dirac delta function).
Since a vortex at $\R$ and a domain wall in $\h$ are related by a global rotation to a configuration with $-\R$ and $-\h$, their effective free energy is the same: $F_{\rm eff}(\R,\h) = F_{\rm eff}(-\R,-\h)$. Additionally, the effective free energy changes sign upon reversing the exchange field, i.e., $F_{\rm eff}(\R,-\h) = - F_{\rm eff}(\R,\h)$; therefore $F_{\rm eff}(-\R,\h) = - F_{\rm eff}(\R,\h)$. This explains why
$F_{\rm eff}^{(nw)}$ and $F_{\rm eff}^{(bw)}$ are odd functions in $\tilde{\R}$, and in particular the second-order contributions vanish in both cases. 
The expansion coefficients in Eqs.~\eqref{eq:FeffNWBW} are determined by the integrals $c_i = \int_{-\tilde{L}_y}^{\tilde{L}_y} \eta_i d\tilde{y}$, where $\tilde{y}= y / \xi$, $\tilde{L}_y= L_y / \xi$, and the dimensionless functions $\eta_i$ are given by the expressions
\begin{align}
\eta_1 =& \frac{1}{ \tilde{y}^2 }\tanh^2 (\tilde{y}) \,, \\
\eta_2 =& \frac{1}{\tilde{y}^4 }\tanh^2 (\tilde{y}) [ 1- \tilde{y} \csch (\tilde{y}) \sech (\tilde{y}) ] \,, \\
\eta_3 =& \frac{1}{\tilde{y}^4} [ ( 3 + 2\tilde{y}^2 + 4\tilde{y}\tanh(\tilde{y}) ) \sech^2 (\tilde{y}) \nonumber \\
&\quad\quad - 3 - 3\tilde{y}^2 \sech^4 (\tilde{y}) ] \,, \\
\eta_4 =& \frac{1}{ \tilde{y}^2 } \tanh (\tilde{y}) [ 2\tilde{y}\sech^2 ( \tilde{y} ) - \tanh (\tilde{y}) ] \,, \\
\eta_5 =& \frac{1}{3\tilde{y}^4}[ ( 3 + 6\tilde{y}^2 + 2\tilde{y}( 3 + 2\tilde{y}^2 ) \tanh(\tilde{y}) ) \sech^2(\tilde{y})  \nonumber\\
&\quad\quad - 3 - 3\tilde{y}^2 (3 + 4\tilde{y}\tanh(\tilde{y}) ) \sech^4(\tilde{y}) ] \,, \\
\eta_6 =& \eta_3 \,. 
\end{align}
While exact analytic expressions for $c_1$ and $c_2$ are not obtainable, the other coefficients read
\begin{align}
c_3 =& \frac{2}{\tilde{L}_y^3} \left[ 1 - \sech^2(\tilde{L}_y) (1 + \tilde{L}_y \tanh^{}(\tilde{L}_y)) \right] \,, \\
c_4 =& \frac{2\tanh^2 (\tilde{L}_y)}{\tilde{L}_y} \,, \\
c_5 =& \frac{\sech^4 (\tilde{L}_y)}{12\tilde{L}_y^3} [\cosh^{}(4\tilde{L}_y) + 16\tilde{L}_y^2 \nonumber \\
& - 1 - 8\tilde{L}_y^2\cosh^{}(2\tilde{L}_y) - 8\tilde{L}_y\sinh^{}(2\tilde{L}_y)] \,, \\
c_6 =& c_3 \,. 
\end{align}
Importantly, all the expansion coefficients are positive for $L_y/\xi > 2$, depend on the system size $L_y$, and approach the values $c_1 \approx 3.41$, $c_2 \approx 1.32$, $c_3 \approx 2(\xi/L_y)^3$, $c_4 \approx 2\xi/L_y$, $c_5 \approx (2/3)(\xi/L_y)^3$, and $c_6 \approx 2(\xi/L_y)^3$ when $L_y/\xi \gg 1$. Thus, the terms proportional to $c_{3-6}$ account for finite-size effects, as these coefficients vanish in the limit $L_y/\xi \rightarrow \infty$.

The expressions in Eqs.~\eqref{eq:FeffNWBW} capture the functional form of the effective free energy for vortices located in proximity to N\'eel and Bloch walls. 
In the case of N\'eel walls, the series expansions of the $F_0^{(1)}$ and $F_0^{(2)}$ terms are identical, resulting in an energy minimum at $\R_0 = -{\rm sign}(\epsilon F_0^{(1)} + F_0^{(2)}) \xi \sqrt{c_1/3c_2}\;\xhat$.
In contrast, for heterostructures containing Bloch walls, the $F_0^{(1)}$ term exhibits an energy minimum at $R_y = -{\rm sign}(\epsilon F_0^{(1)}) L_y$. 
Note that this phenomenon is solely a finite-size effect, as it stems from the $c_{4-6}$ terms in $F_{\rm eff}^{(bw)}$ that vanish as $L_y/\xi$ tends to infinity. 
The $F_0^{(2)}$ term, on the other hand, still has a minimum at $\R_0 = -{\rm sign}(F_0^{(2)}) \xi \sqrt{c_1/3c_2}\;\xhat$.

The main features of the numerically computed effective free energies in Fig.~\ref{fig:Fme} of the main text follow from Eqs.~\eqref{eq:FeffNWBW} after setting $F_0^{(2)} = 0$. This is equivalent to a vanishing $\kappa^{(2)}$; whose contribution is negligible when compared to $\kappa^{(1)}$, as we justify in App.~\ref{app:js_plots-kappa2}.
Furthermore, our numerical simulations are consistent with the $F_0^{(1)}$-term in Eqs.~\eqref{eq:FeffNWBW}.
For a N\'eel wall with $\gamma= 0$, the $F_0^{(1)}$-term exhibits a minimum at $\R= - \xi \sqrt{c_1/3c_2}\;\xhat$. This explains why the vortex crosses into the left domain and is dragged behind the moving wall. 
Conversely, for a N\'eel wall with $\gamma= \pi$, the vortex reaches equilibrium at $\R= \xi \sqrt{c_1/3c_2}\;\xhat$ and is pushed in front of the domain wall as it moves to the right.
In the case of Bloch walls, the $F_0^{(1)}$-term has a minimum at the end of the wall, given by $R_y= L_y$ ($R_y= -L_y$) for $\gamma = 3\pi/2$ ($\gamma = \pi/2$). 
This effective free energy landscape gives rise to a $y$-component in the vortex's motion, which explains its gliding movement along the wall.

\section{Supercurrent density and second-order SOC term significance}\label{app:js_plots-kappa2}

The vector fields of the supercurrent density $\j_s$ plotted in Fig.~\ref{fig:DW-SC} in the main text were numerically computed using the eigenenergies and eigenstates of the tight-binding Hamiltonian in Eq.~\eqref{eq:TightBindingHamiltonian} of the main text written in Bogoliubov-de Gennes form. 
For this numerical calculation of $\j_s$ we used the lattice expression from the appendix of Ref.~\cite{Hals2016b}. Comparing the vector fields of $\j_s$ in Fig.~\ref{fig:DW-SC} with the prediction from Ginzburg-Landau theory in App.~\ref{app:GL_theory}, reveals that the term in $\j_{as}$, Eq.~\eqref{eq:app_jsA}, proportional to $\kappa^{(2)}$ does not contribute significantly.

To assess the significance of the term proportional to $\kappa^{(2)}$ (also $F_0^{(2)}$ in Eqs.~\eqref{eq:FeffNWBW}), we performed a numerical analysis of the relative magnitude 
of $j_{as, x}^{(dw)} $ and $j_{as, y}^{(dw)} $ in Eq.~\eqref{eq:app_jsDW} for a system containing a Bloch wall without a vortex.
In this case, the $\kappa^{(1)}$-term generates a current along the $x$-axis, while the $\kappa^{(2)}$-term produces a current parallel to the $y$-axis.
Consequently, the ratio $j_{as, y}^{(dw)} / j_{as, x}^{(dw)} $ provides a direct measure of $\kappa^{(2)}/\kappa^{(1)}$.
Our numerical investigation shows that the value of $j_{as, y}^{(dw)} / j_{as, x}^{(dw)} $ is significantly below the machine precision, implying that the contribution from $\kappa^{(2)}$ is negligible.


%

\end{document}